\documentclass[twocolumn, superscriptaddress,prl]{revtex4-1} 
\usepackage{graphicx} \usepackage{epsfig} \usepackage{color}
\usepackage{braket}
\usepackage{amsmath,amsfonts,amsthm,bm}
\usepackage[utf8]{inputenc}
\usepackage[T1]{fontenc}

\usepackage[normalem]{ulem}

\usepackage[breaklinks,colorlinks,
linkcolor=blue,citecolor=blue,urlcolor=blue]{hyperref}
\begin{document}
\title{
Emergent periodic and quasiperiodic lattices on surfaces of 
synthetic Hall tori and synthetic Hall cylinders 
}
\author{Yangqian Yan}
\affiliation{Department of Physics and
Astronomy, Purdue University, West Lafayette, IN, 47907}
\author{Shao-Liang Zhang}
\affiliation{School of Physics, Huazhong University of Science and Technology, Wuhan 430074, People's Republic of China}
\author{Sayan Choudhury}
\affiliation{Department of Physics and
Astronomy, Purdue University, West Lafayette, IN, 47907}
\author{Qi Zhou}
\affiliation{Department of Physics and
Astronomy, Purdue University, West Lafayette, IN, 47907}
\date{\today}
\begin{abstract}

Synthetic spaces allow physicists to bypass constraints imposed by certain
physical laws in experiments. Here, we show that a synthetic torus, 
which consists of a ring trap in the real space and internal states of 
ultracold atoms cyclically coupled by Laguerre-Gaussian Raman beams,
could be threaded by a net effective magnetic flux through its surface---an
impossible mission in the real space. Such synthetic Hall torus gives rise to 
a periodic lattice in the real dimension, in which the periodicity of density 
modulation of atoms fractionalizes that of the Hamiltonian.  Correspondingly, 
the energy spectrum is featured by multiple bands grouping into clusters with 
nonsymmorphic-symmetry-protected band crossings in each cluster, leading to 
swaps
of wavepackets in Bloch oscillations. Our scheme allows physicists 
to glue two synthetic Hall tori such that localization may emerge in a 
quasicrystalline lattice. If the Laguerre-Gaussian Raman beams and ring traps 
were replaced by linear Raman beams and ordinary traps, a synthetic Hall 
cylinder could be realized and deliver many of the aforementioned phenomena.

\end{abstract}

\maketitle

Spaces with nontrivial topologies provide quantum systems 
unprecedented 
properties~\cite{Kleinert1989,Schulte1997,Kaneda2001,Lu2001,Turner2010,Ho2015a,Guenther2017}.
As a prototypical
space of a finite genus, the importance of a torus in modern physics is more
far-reaching than applying periodic boundary conditions (PBC) in
theoretical 
calculations. 
It plays a crucial role in quantum Hall physics. 
The ground state of a fractional quantum Hall state becomes degenerate on a torus or any surface with a finite genus~\cite{Wen1990}.
Such degeneracy, which is unavailable on a cylinder or a flat space and defines the concept of the topological order
lays the foundation of topological quantum computation~\cite{Kitaev2003}.
However, due to the absence of magnetic monopoles in nature,
it is impossible to generate a net magnetic flux through a closed surface
in the real space.
The study of  quantum Hall states on a torus has eluded experiments so
far.

Ultracold atoms provide physicists 
a unique platform to 
engineer 
Hamiltonians and allow physicists to achieve many quantum Hall states unattainable in electronic systems~\cite{Bloch2008},
such as quantum Hall states of bosons and quantum Hall states with high
spins.
Other than the typical harmonic potentials,
ring traps have been implemented 
in an annular
geometry\cite{Morizot2006,Eckel2014}. 
Linear and Laguerre-Gaussian (LG)
Raman beams have been used to create spin-momentum coupling and spin-angular
momentum coupling, respectively.~\cite{Lin2011,Sun2015,Chen2016,Chen2018,Zhang2018}. 
If one considers the internal degree of freedom
as a synthetic dimension, the spin-momentum coupling gives rise to a 
synthetic magnetic field in a two-dimensional plane~\cite{Celi2014,Anisimovas2016}. Whereas experiments have 
been focusing on open boundary conditions in the synthetic dimension~\cite{Stuhl2015,Mancini2015},
there have been 
theoretical proposals on creating a periodic or twisted
boundary 
condition~\cite{Grusdt2014,Boada2015,Acki2016,Budich2017,Taddia2017,Kim2018a}.
However, few experiments has fulfilled the requirements of 
these proposals~\cite{Han2018,Li2018}.

We propose a simple scheme to realize a synthetic torus penetrated by a
net effective magnetic flux.  Ultracold atoms confined in a ring trap 
in the real space 
are subjected to spin-angular momentum coupling induced by LG 
Raman beams.
Either hyperfine spins or nuclear spins could be used to enable a cyclic
coupling and
form a loop in the discrete synthetic dimension. 
Cyclic couplings have been studied 
for different purposes, including realizing two-dimensional spin-orbit coupling and creating Yang monopoles~\cite{Campbell2011,Huang2016,Sugawa2018}.
Here, we use spin-angular momentum coupling to synthesize internal states and the real dimension into a synthetic Hall torus. 
PBCs in both the synthetic and the real dimension
deliver a 
torus. Spin-angular momentum coupling produces
 finite effective magnetic fluxes penetrating its toroidal surface,
signifying the rise of a synthetic Hall torus.
Replacing Laguerre-Gaussian Raman beams by linear ones,
our scheme applies to ordinary traps with open boundary conditions for
creating synthetic Hall cylinders, which have been realized by 
experiments recently~\cite{Han2018,Li2018}. 

We further unfold 
unique properties
of 
synthetic 
Hall tori and cylinders.
Unlike previous works including optical lattices in the real
dimension~\cite{Stuhl2015,Mancini2015,Acki2016,Taddia2017,Kim2018a,Han2018}, we consider a continuous
real space trap. Interestingly, 
periodic or quasiperiodic 
lattices emerge in the continuous real dimension,
as a result of PBC in the synthetic dimension. 
The periodic lattice modulates the density of atoms with a fractionalized
periodicity of the Hamiltonian 
and a unique band structure shows up. 
Energy bands form clusters with nonsymmorphic-symmetry-protected band 
crossings in each cluster. Wavepackets in each cluster swap with each other in Bloch 
oscillations. 
Though
each single synthetic Hall torus or cylinder supports only extend states, once
two of them are glued together, quasiperiodic lattices may emerge and lead to
localized states in the real space. Such ``localization from gluing"
demonstrates the power of synthetic Hall tori or cylinders in accessing
even more complex synthetic spaces and intriguing quantum phenomena there. 

{\it Proposed scheme and Hamiltonian.}
We consider $M$ 
internal states in a real space ring trap.
For alkali atoms, 
these $M$ spins involves both $F=1$ and $F=2$, as shown in Fig.~\ref{fig1}(a).  
\begin{figure} \centering
  \includegraphics[angle=0,width=0.45\textwidth]{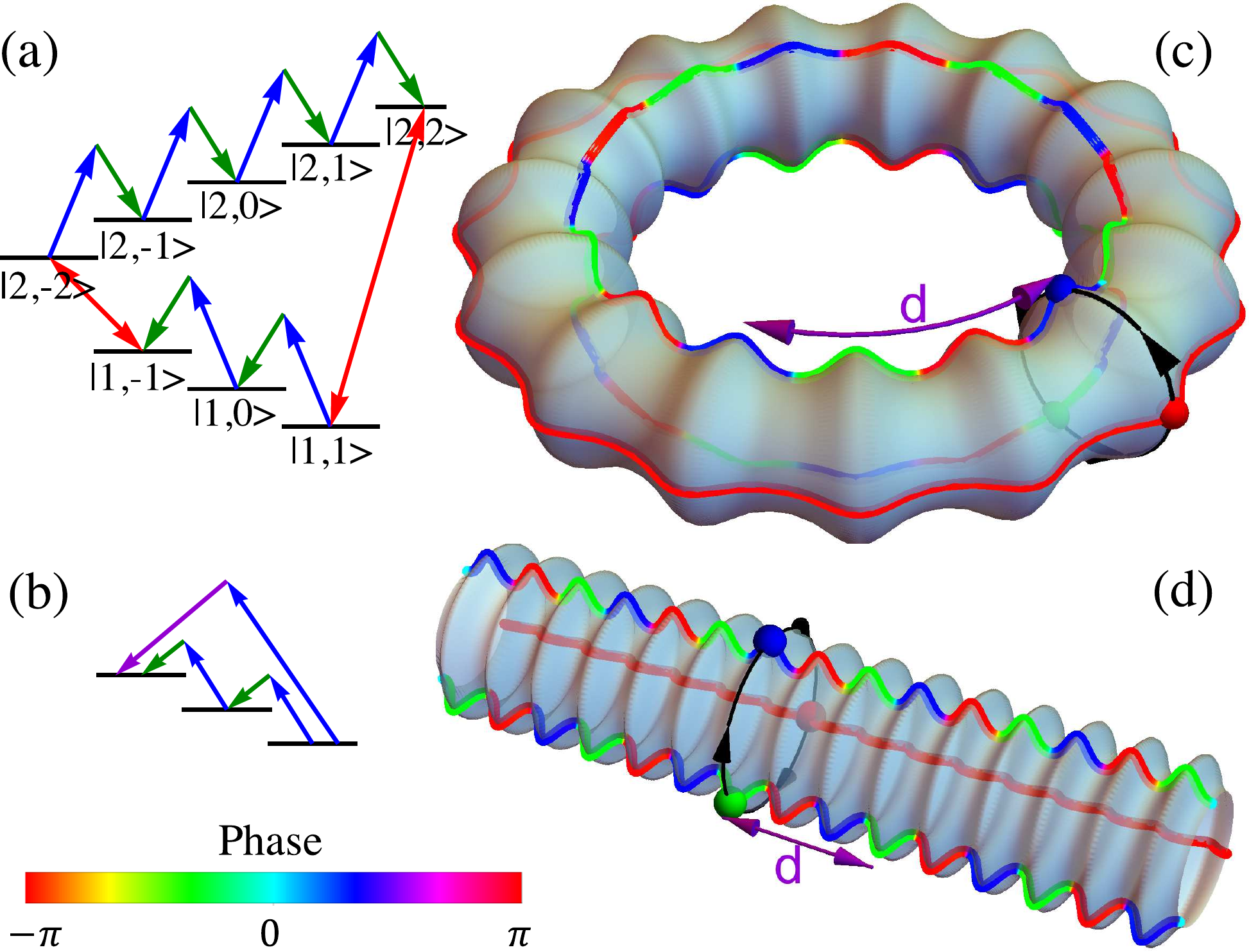}
  \caption{(Color online) 
    (a) Energy diagram for the hyperfine states and the laser coupling scheme.
    Blue and green arrows
    represent the Raman coupling. Bidirectional arrows represent
    microwave couplings.
    (b) Simplified coupling diagram for
    three
    internal 
    states coupled by Raman
    beams.
    (c-d) Torus (Cylinder) formed by cyclically
    coupled three internal states
    in a real space ring trap.
    The density oscillation is depicted as the fluctuation of the radius of the torus or
    cylinder. 
    Colored curves represent phases of the wavefunction of each spin
component.  } \label{fig1} \end{figure}
At weak magnetic fields, linear Zeeman splitting dominates, thus a single pair of LG
Raman beams simultaneously couples every consecutive states within each manifold; Microwave
fields couple $|1,1\rangle$ ($|1,-1\rangle$) and $|2,2\rangle$
($|2,-2\rangle$). These eight states 
form a circle in the synthetic dimension. Due to the opposite $g$ factors
between $F=1$ and $F=2$, 
a finite angular momentum
transfer occurs once an atom finishes the loop in the synthetic dimension. 
A net effective magnetic flux 
emerges on the torus.
Each hyperfine spin state has multiple angular momenta.

The number of internal states are 
controllable. 
On the one hand,
at large magnetic fields, quadratic Zeeman
splittings become important, and fewer spins 
can be separated out from
the rest to form a smaller circle~\cite{Li2018,Campbell2011,Sugawa2018,Taddia2017,Han2018}.  
  Note that shrinking the synthetic dimension does not change any results qualitatively,
  as fractional quantum Hall states can be adiabatically connected to 1D charge-density waves~\cite{Seidel2006}.
  Chiral edge currents in the quantum Hall strips have also been observed using only three
internal states~\cite{Stuhl2015,Mancini2015}.
On the other hand,
using the $^1S_0$ and $^3P_0$ states of Sr87~\cite{Boyd2006}, one could cyclically couple
up to 20 internal states.

Here, we consider nearest neighbors in the synthetic dimension coupled by LG
Raman beams. The spin flip from the $j$th spin state to the $j+1$th one is
thus accompanied by an angular momentum increase $m_{j,j+1}$, 
which 
is the difference between the angular momenta carried by the two
LG beams.
A microwave coupling then corresponds to $m_{j,j+1}=0$.

We define the position $x=\phi L/(2\pi)$
and the momentum $p=2\pi m/L$,
where $\phi$ is the azimuthal angle and $L$ is the circumference of the real
space ring. 
We also define $q_{j,j+1}=2\pi m_{j,j+1} /L$ as
the ``momentum'' transfer along the azimuthal direction. 
The advantage of the notation is that all results directly apply to a cylinder.  
 In both the cylinder and the torus, $x$ represents the direction in the real dimension.
The Hamiltonian reads
\begin{equation}
\begin{split} H&=\sum_{j=1}^M
|\psi^{j}(x)\rangle(-\frac{\hbar^2}{2m_0}\partial^2_x +\epsilon_j)\langle
\psi^j(x)| \\ +&\sum_{j=1}^M
\left(\Omega_{j,j+1}e^{iq_{j,j+1}x}|\psi^{j+1}(x)\rangle\langle
\psi^j(x)|+h.c.\right),\label{H} \end{split} \end{equation}
where  $\psi^j(x)$ denote the spacial wave function for the $j$th spin,
$\epsilon_j$ the one (two) photon detuning in the microwave (Raman)
transition, $\Omega_{j,j+1}$
the coupling strength between the $j$th and the $j+1$th spin state, and $\psi^{M+1}(x)=\psi^{1}(x)$. 
Whereas our results are very general and do not require tuning every single parameter arbitrarily,
quadratic Zeeman splitting could create uneven energy separations and
each pair of hyperfine spin states could be coupled by different lasers.
Thus, $\Omega_{j,j+1}$ can be, in principle, tuned independently.
The total phase
 accumulated after an atom finishes a circle $j\rightarrow
 j+1\rightarrow j+2... \rightarrow j-1\rightarrow j$,
 $\varphi(x)\equiv e^{i Q x}$ is finite and spatially dependent, where
 $Q\equiv\sum_{j=1}^M q_{j,j+1}$.
 The total synthetic magnetic flux on the surface of the torus per unit
 length in the physical dimension is then propotional to $Q$.  

 \textit{Nonsymmorphic symmetry and band structures.} We start from commensurate
momentum transfers, i.e.,  $q_{j,j+1}=n_j q_L$, where $n_j$ are integers.
For any coupling strengths $\Omega_{j,j+1}$,
the reciprocal lattice vector $q_L$
determines the periodicity of $H(x)$,  $H(x)=H(x+\frac{2\pi}{q_L})$. If one of these couplings vanishes, the phases, $e^{iq_{j,j+1}}$,
can be absorbed to $|\psi^{j+1}(x)\rangle$, and any spin component in an eigenstate contains a single plane wave. 
In contrast, when all $\Omega_{j,j+1}\neq 0$, the phases cannot be gauged away under such PBC.
A single spin component in any eigenstate contains multiple plane waves and the densities form standing waves.
The lattice in the real space is therefore an emergent one from the PBC in the synthetic dimension.

The Bloch wavefunctions $\vec{\psi}_k(x)$ are simultaneous eigenstates of $H$
and $\hat{T}(d)$ [$\hat{T}\left( d \right)\vec{\psi}_k(x)=e^{ikd}\vec{\psi}_k(x)$],  where $\vec{\psi}(x)$
 is a $M$-component wavefunction, $k$ the quasimomentum, and $d\equiv
\frac{2\pi}{q_L}$ the lattice spacing, $T(d)$ the translation operator of
distance $d$.
We define the nonsymmorphic symmetry operator $\hat{G}$ as a combination of a translation for a fraction of the lattice
space $T(2\pi/Q)$ in the real dimension and a unitary transformation $U_s$ in the synthetic
direction, 
\begin{equation} 
  x\rightarrow x+\frac{2\pi}{Q},\,\,\, |\psi^{j>1}\rangle\rightarrow
  e^{-i \frac{2\pi}{Q}\sum_{j'=1}^{j-1} q_{j',j'+1}} |\psi^j\rangle.
\end {equation}
Again, it is understood that $M+1$ is equivalent to $1$.
A simple example of the nonsymmorphic symmetry is the glide-reflection symmetry,
a translation for half of the lattice spacing combined with a reflection in the perpendicular direction,
which has played an important role in topological quantum matters~\cite{Shiozaki2015,Fang2015,Wang2016}.
Consider
a special case,
$\epsilon_j=0$, $\Omega_{j,j+1}=\bar{\Omega}$ and $n_j=\bar{n}$, the synthetic
dimension becomes translational invariant in the real dimension. $\hat{G}$ and its multiples, together
with the translation in the synthetic dimension, then form the conventional
magnetic translation group~\cite{Zak1964}.  
In generic cases where the
synthetic dimension does not have translation invariance, i.e., 
nonuniform $\epsilon_j$, $\Omega_{j,j+1}$ or $n_j$. $[H,
\hat{G}]=0$ is still satisfied and signifies a nonsymmorphic symmetry.

We define $n\equiv Q/q_L=\sum_{j=1}^M n_j$.
Many physical quantities depend on $n$, i.e., properties of
the system crucially rely on how the synthetic magnetic field is distributed
on the surface of the torus, not just the total flux. Applying $\hat{G}$ for $n$
times is equivalent to a translation in the physical dimension for one
lattice spacing, $\hat{G}^{n}\vec{\psi}_k(x)=e^{ikd}\vec{\psi}_k(x)$. Thus,
\begin{equation}
  \hat{G}\vec{\psi}_{k}(x)=c_s\vec{\psi}_k(x),\,\,\,\, c_s=e^{i(\frac{kd}{n}+\frac{2 s\pi}{n})},
\label{geigen}
\end{equation}
where $s=1,2,... n-1,n$. Equation~(\ref{geigen}) shows that, as the quasimomentum $k$
changes by a reciprocal lattice vector $q_L$, the eigenvalue of $\hat{G}$ changes by
$e^{2i\pi/n}$, i.e., the $s$th eigenvalue becomes the $s+1$ one.  Meanwhile,
$\vec{\psi}_k(x)=\vec{\psi}_{k+q}(x)$ is satisfied. Thus, we conclude that bands must form
clusters, each of which contains $n$ bands. These $n$ bands are the $n$
eigenstates of the operator $\hat{G}$ with the $s$th eigenvalue $c_s$, and
intersect with each other within the Brillouin zone (BZ). 

We solve $H$ in Eq.~(\ref{H}) 
using plane-wave expansions. 
The band structure fully agrees with the prediction from the above symmetry considerations.
Fig.~\ref{fig2}(a)
\begin{figure} \centering
  \includegraphics[angle=0,width=0.45\textwidth]{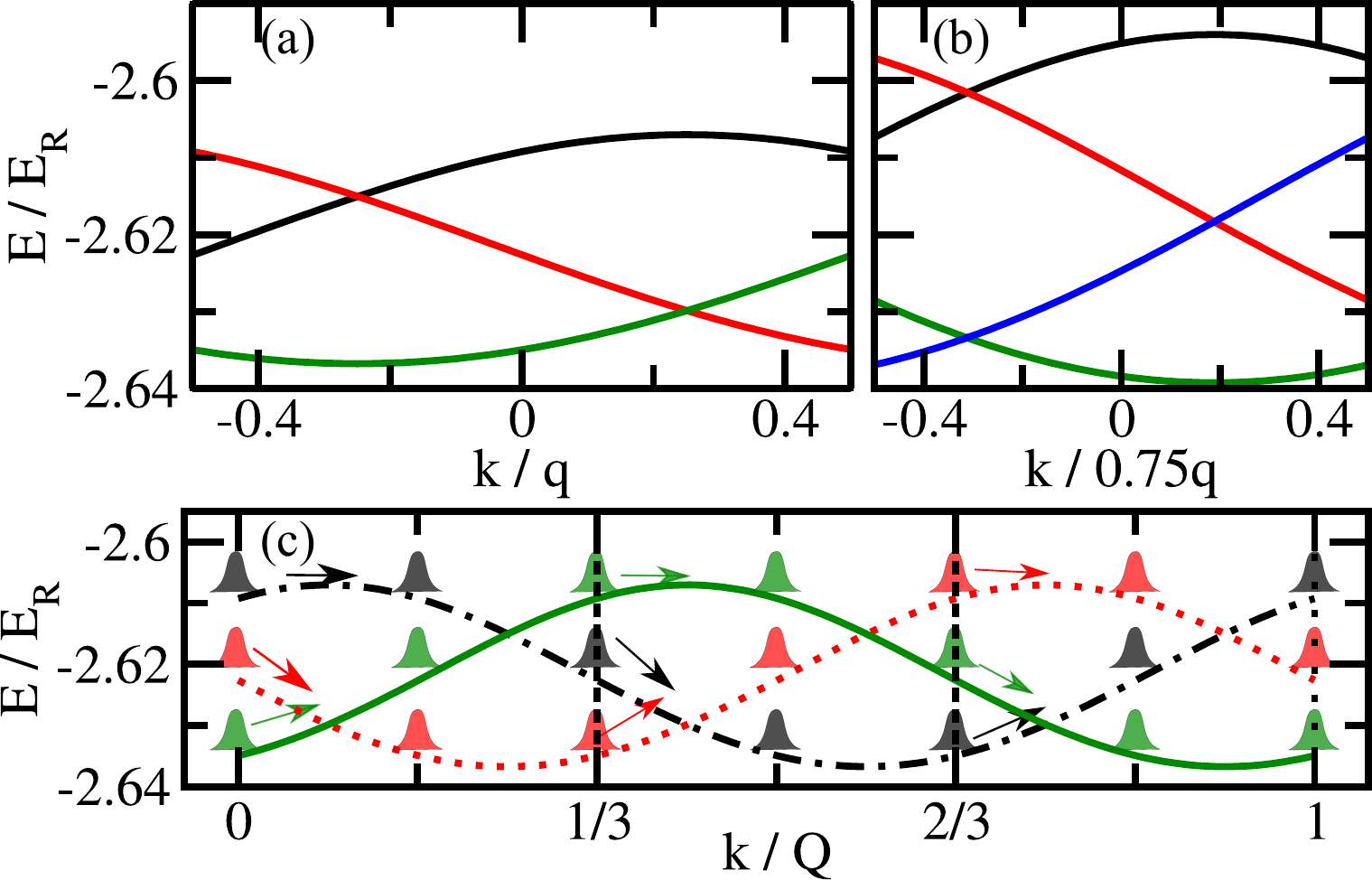}
  \caption{ 
    Band structures when $q_{1,2}:q_{2,3}:q_{3,1}=1:1:1$ (a) and $q_{1,2}:q_{2,3}:q_{3,1}=1:1:2$ (b). 
    (c): Bloch oscillation for (a). Vertical lines represent the boundaries of BZ.
  } \label{fig2} \end{figure}
shows the energy bands when $M=3$,  
$q_{1,2}=q_{2,3}=q_{3,1}=q$, and coupling strength $\Omega_{1,2}=1.2E_R$, $\Omega_{2,3}=1.8 E_R$, and
    $\Omega_{3,1}=1.5E_R$, where $E_R=\hbar^2Q^2/2m_0$ is the recoil energy
    defined by $Q$. Here, the reciprocal lattice vector $q_L=q$ and $n=Q/q_L=3$. 
   The cylindrical or toroidal surface is penetrated by a uniform flux.
    Thus,
three bands exist 
in each
cluster. The eigenstate 
of the $s$th band 
is 
\begin{eqnarray}
  \vec{\psi}_k(x)&=&e^{i (k+sq) x}(u^1_{k}(x), u^2_{k}(x),u^3_{k}(x))^T, \label{wf1}\nonumber\\ 
u^j_{k}(x)&=&e^{i (j-1)q x}\sum_{l=-\infty}^{\infty} c^j_{l}(k)e^{i  l Q x},
\label{wf2} \end{eqnarray}
where $u^j_{k}(x)$ is the periodic Bloch wavefunction of the $j$th spin state,
$l$ an integer, and $c_{l}^{j}(k)$ determined by Eq.~(\ref{H}).
The density of the $j$th spin state, $\rho^j_{k}(x)\equiv |u^j_{k}(x)|^2$
satisfies
\begin{equation}
\rho^j_{k}(x)=\rho^j_{k}(x+\frac{2\pi}{Q})=\rho^j_{k}(x+\frac{d}{3}).
\label{periodeq}
\end{equation}
The total density $\rho(x)=\sum_j\rho^j_{k}(x)$, by definition, also 
satisfy Eq.~(\ref{periodeq})~\cite{note1}.
Despite the continuous real dimension, the density of atoms
oscillates with a period only $1/3$ of that of the Hamiltonian,
as shown in Fig.~\ref{fig1}(c-d).
In contrast, the relative phase between $u^2_{k}(x)$ [$u^3_{k}(x)$] and
$u^1_{k}(x)$ has a periodicity of $d$, as shown by the colored curves in
Fig.~\ref{fig1}(c-d).
As previously discussed, such results
crucially depend on the PBC. 
Both periodic density and phase oscillations vanish once the synthetic dimension  has an open boundary condition. 

{\it Swapping wave packets in Bloch oscillations}. 
When a constant force is applied,  a
wavepacket in the momentum space 
experiences a Bloch oscillation, which has
exactly the same period of the Hamiltonian in an ordinary band structure. In contrast, the period of the
Bloch oscillation
here
is given by $3q$, tripling the reciprocal lattice
vector. Due to the presence of band crossings, a wavepacket does not
return to the original band after the momentum changes by $q$.  Instead, it swaps with another wavepacket from a different band.  
 
 {The significance of such Bloch oscillations here is that it traces}  the
Wilson lines, a path-ordered integral of non-abeliean Berry connections in the
momentum space~\cite{Li2016,Zhang2017}. Due to the nonsymmorphic-symmetry-protected band crossings,
abelian Berry connections no longer applies when studying topological properties of the band structure.
Non-abelian Berry connections and the Wilson lines characterize how a quantum state changes to a different
one while the Hamiltonian returns to the original one~\cite{Wilczek1984,Fruchart2019}, a prototypical non-abelian operation.
This is precisely what we see from Fig.~\ref{fig2}(c). If we label the bands as $1, 2, 3$ from bottom to top based on the energies,
the green wave packet initially at band-1 moves to band-3, meanwhile the red (black) one initially at band-2 (band-3) moves to band-1 (band-2) when
 $\Delta k=q$~\cite{note2}.
When $\Delta k=Q=3q$, these three wave packets swap with each other for six times, 
as shown in Fig.~\ref{fig2}(c).
 
The above discussions can be directly generalized to other choices of $\{q_{j,j+1}\}$.
If $q_{1,2}=q_{2,3}=\frac{3}{4}q$, $q_{3,1}=\frac{3}{2}q$, though the total momentum transfered, $Q$, is still $3q$, the same as the previously discussed case, $q_L$ becomes $3q/4 $ and $n=4$. One thirds of the surface has a
larger magnetic flux than the remaining region~\cite{note2}, and a cluster consists of four
bands [Fig.~\ref{fig2}(b)]. Changing the value of some of the wavevectors is equivalent to
redistributing the magnetic flux on the surface, and leads to distinct band
structures. We emphasize that the total number of
states of the system remains unchanged. The change of the number of bands is
associated with the change of BZ. In this example,
the reciprocal lattice vector $q_L'$ become $\frac{3}{4}q$.
Shrinking the size of BZ then leads to an increase of the number of
bands in a cluster. 

{\it Quasiperiodic lattices}. If $q_{j,j+1}$ are incommensurate, e.g.,
$q_{1,2}:q_{2,3}:...:q_{M,1}=1:1:...:\gamma$, where $\gamma=\frac{\sqrt{5}-1}{2}$, a peculiar quasiperiodic
lattice arises: certain quantities have well defined
periodicities but others do not. For example, 
Eq.~(\ref{wf1}-\ref{wf2}) holds for a generic $\{q_{j,j+1}\}$ when $M=3$.
Though $H$ in Eq.~(\ref{H}) is aperiodic, the density of each
spin still satisfies $\rho^j_{k}(x)=\rho^j_{k}(x+\frac{2\pi}{Q})$.
The wavefunction of each spin component is still extended, as its plane wave
expansion only includes multiples of $Q$. In contrast, the relative phases
between different spin component are spatially variant and are not
commensurate.
Thus, the wavefunction $\psi_k(x)$ is aperiodic in the real
dimension. Defining a pseudospin-1,
$S_\mu=\sum_{j,j'}u^{j*}_k(x)\mathcal{F}_\mu^{j,j'}u^{j'}_k(x) $, where
$j,j'=1,2,3$,
$\mathcal{F}_\mu^{j,j'}$ are the spin-1 Pauli matrices, and $\mu=x,y,z$.
$S_z(x)$ is periodic, $S_z(x)=S_z(x+2\pi/Q)$, but $S_x(x)$ and $S_y(x)$ do not have well
defined periods, as shown in Fig.~\ref{fig34}(a). 
\begin{figure} \centering
  \includegraphics[angle=0,width=0.45\textwidth]{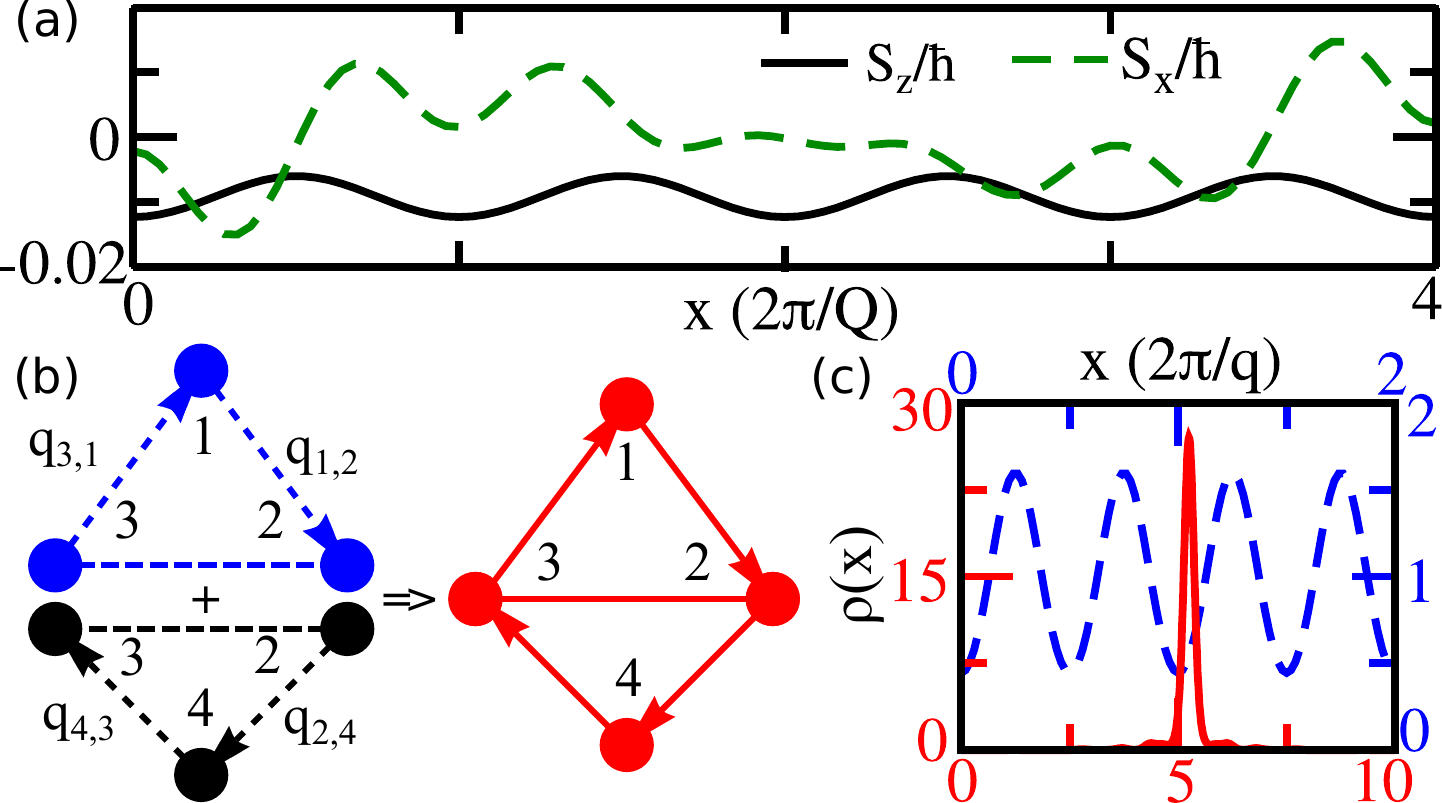}
  \caption{(a) Spin polarization along the $x$ and $z$
    direction as a function of $x$.
    (b) 
    Schematic of the cross sections of two tori or cylinders when they are
    glued together. 
    (c) Total density of the ground eigenstate at zero quasimomentum 
    before (dashed line, only the top blue torus or cylinder is shown) and after (solid line) the gluing.
    $q_{1,2}=q_{3,1}=q$, $q_{2,3}=0$, $q_{2,4}=q_{4,3}=\gamma q$,
    and all couplings $\Omega_{j,j'}$ are $2E_r$,
    where $E_r=\hbar^2 q^2/2m_0$ is the
    recoil energy defined by $q$.
    $\gamma_8=13/21$ has been used as as an approximation
    of $\gamma=(\sqrt{5}-1)/2$. 
}
  \label{fig34} \end{figure}

On a cylinder, there is no restriction on the choice of $q_{j,j+1}$. In
contrast, 
an irrational ratio $q_{j,j+1}/q_{j',j'+1}$ is
not allowed on a torus, as the PBC in the real
dimension require that all momentum scales are multiples of $2\pi/L$.
Nevertheless, 
any irrational number can be approached by the
ratio of two integers with increasing the integers' values. For instance,
$\gamma$ can be approximated
by $\gamma_\alpha=a_{\alpha-1}/a_{\alpha}$, where $\{a_\alpha\}$ is the
Fibonacci series 1,1,2,3,5\dots, with increased
accuracy.
When the approximation order $\alpha$ increases, 
the periodicity of $S_{x,y}(x)$
increases. 
$\gamma_\alpha$ with a small $\alpha$ well reproduces the result for a small $x$ and a large $\gamma$.

{\it Localization by gluing.}
Our scheme can be implemented to access more complex synthetic spaces. For instance, 
adding extra couplings to the synthetic dimension is equivalent to gluing multiple tori or
cylinders. Figure~\ref{fig34}(b) shows that two tori or cylinders with $M=3$
can be glued to a single one with $M=4$. In the real space, it is difficult to realize such gluing, as it is required to identify certain parts of two different objects. Using the synthetic dimension, adding an additional tunneling
through the interior of single tori or cylinder with $M=4$, $\Omega'=\Omega_{2,3}$, immediately realizes this gluing and delivers a system with a different topology. Though each single torus or
cylinder supports only extended eigenstates, after gluing them together,
eigenstates at low energies could become localized, as shown in
Fig.~\ref{fig34}(c)~\cite{note3}.
The wavefunction of a
single spin component now includes multiple momentum scales, $q_{1,2}$,
and $q_{4,3}$, unlike a single torus or cylinder case where only $Q$
is relevant. The interference of plane waves with incommensurate wave vectors
could thus potentially localize the wavefunction.

To quantitatively characterizes the localization, we compute the width of the
lowest band as a function of $\Omega'$. 
It has been shown that the ground band width scales with $a_{\alpha}^{-2}$ for extended states, 
and decays much faster for localized states~\cite{HIRAMOTO1992, Diener2001}. 
Here, the scaled band width almost vanish at an intermediate value of $\Omega'$, where eigenstates are localized, as shown in Fig.~\ref{fig5}.
\begin{figure} \centering
  \includegraphics[angle=0,width=0.45\textwidth]{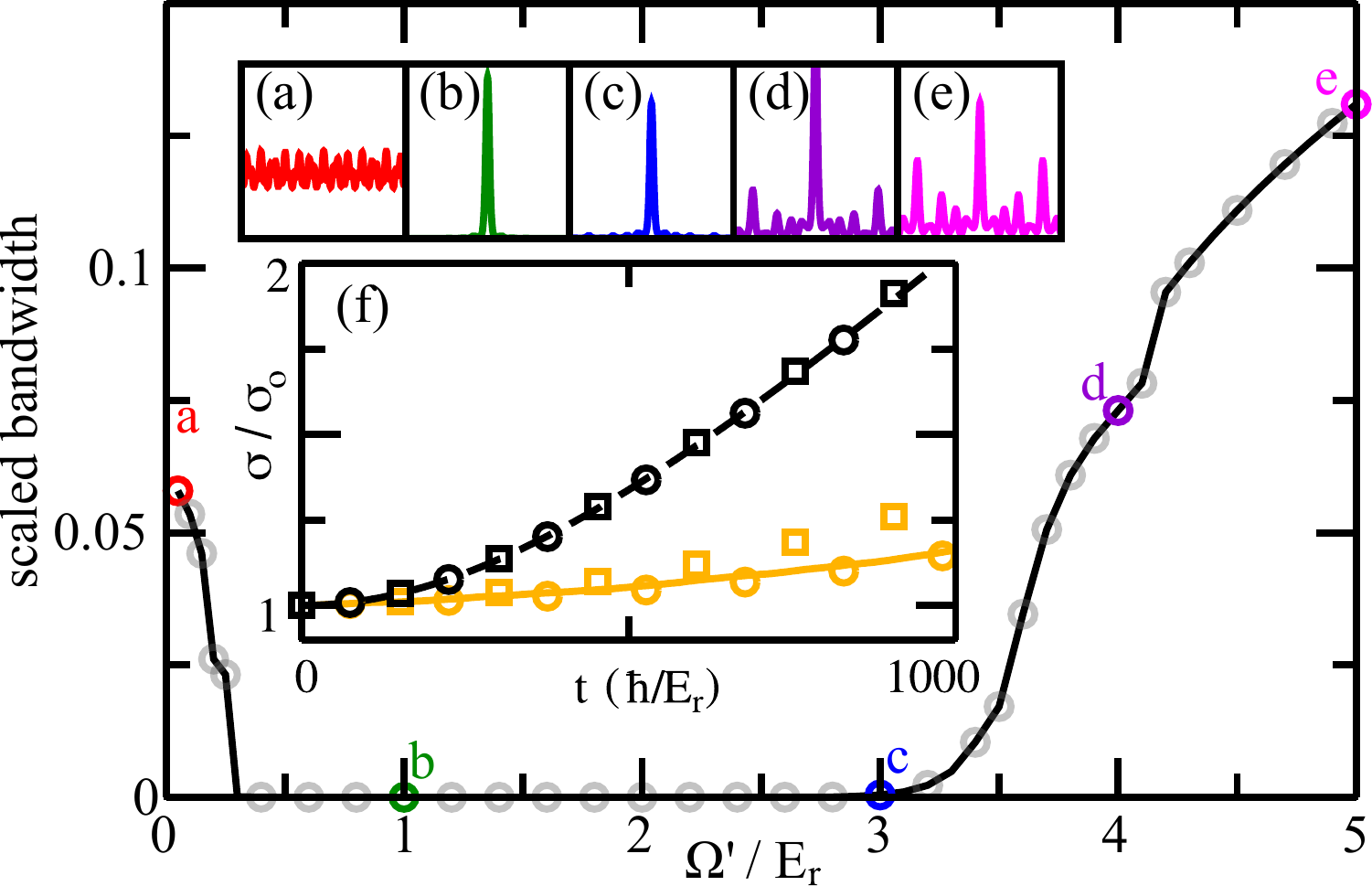}
  \caption{ 
    Bandwidth multiplied by $a_{\alpha}^2/E_r$ as a function of
   $\Omega'$ for 
   $\alpha=12$.
   The
   inset (a),\dots,(e) show the corresponding zero quasi-momentum ground state
   at $\Omega'/E_r=0.05, 1, 3, 4, 5$, respectively with  $\alpha=8$. 
   (f) The 
   width of the Gaussian wave packet as a function of time 
   The solid orange (black dashed) curve represents non-interacting results for $\Omega'=2E_r$ ($\Omega'=0$).
   Squares and circles represent results for $\int g \rho_i^2 \text{d} x=\pm0.003 E_r$, respectively,
   where $\rho_i$ is the density for the initial state.  } \label{fig5} \end{figure}
When $\Omega'$ is very small, the wavefunction is still extended, similar to a single torus without $\Omega'$. 
For large $\Omega'$, dominating contributions to the wavefunction come
from only two hyperfine spin states [2 and 3 in Fig.~\ref{fig34}(b)] such that the incommensurate wave vectors
are no longer relevant and the wavefunction is still extended.

The localization can also be characterized by the expansion of an initially
localized wavepacket with a width $\sigma_0$ in the real space.
We consider a 
Gaussian wave packet as the initial state.
For small or large $\Omega'$, 
where the eigenstates at low energies are delocalized, 
the width of the wave packet, $\sigma$,
increases quickly. In
contrast, $\sigma$ grows much slower in the localized regime. 
To further consider interaction effects, 
we numerically solve a time-dependent Gross-Pitaevskii equation,
\begin{equation} 
i\hbar\frac{\partial\vec{\psi}(x)}{\partial t}= (\hat{H}+ g \rho)\vec{\psi}(x),
\end{equation}
where $g$ is the
interaction strength.
We find that a weak repulsive (attractive) interaction slightly enhance (suppress) the dynamics.
In addition,
we have considered the ground state of interacting bosons in the trap. The main conclusions remain unchanged~\cite{note2}.

We have shown that 
atom-laser interactions allow physicists to
synthesize Hall tori and cylinders 
hosting 
intriguing
quantum phenomena in the emergent periodic and quasiperiodic lattices. We hope
that our 
results 
will stimulate more works on synthetic spaces so as to explore
physics that are not easy to access in conventional traps.

\begin{acknowledgments}
This work is supported by startup funds from Purdue University. 
Support by the National Science Foundation (NSF) through Grant No. PHY-1806796 is gratefully acknowledged.
\end{acknowledgments}


%\end{document}
\clearpage
\pagebreak
\newpage
\widetext
\begin{center}
\textbf{\large 
Supplemental Material of ``Emergent periodic and quasiperiodic lattices on surfaces of 
synthetic Hall tori and synthetic Hall cylinders''
}
\end{center}
\renewcommand{\theequation}{S\arabic{equation}}
\renewcommand{\thefigure}{S\arabic{figure}}
\renewcommand{\thetable}{S\arabic{table}}
\renewcommand{\theHequation}{Supplement.\theequation}
\renewcommand{\theHtable}{Supplement.\thetable}
\renewcommand{\theHfigure}{Supplement.\thefigure}
\setcounter{table}{0}
\setcounter{figure}{0}
\setcounter{equation}{0}

\maketitle
\onecolumngrid

The notation employed in this Supplemental Material follows that introduced in the main text.
Section I used the non-abelian Wilson line technique to clarify why the
crossing is avoided in Fig. 2 of the main text.
Section II clarifies the parameters used in the simulation in Fig. 4 of the
main text.
Section III provides an alternative simulation: expansion of the ground state in a 1D harmonic trap.
Section IV provides a full 3D simulation of a proposed experiment to
demonstrate the localization of the ground state due to gluing of two
cylinders.
Section V clarifies Fig. 2(a-b) in the main text.

\section{ Bloch oscillations in the presence of band crossings}

 In the presence of a constant force, $F$, 
 the quasi-momentum becomes $k_f=k_0+F t$ 
 at time $t$, where $k_0$ is the initial quasi-momentum at $t=0$. 
 In the adiabatic limit, the inter-band transition can be determined by the 
path-ordered integral~\cite{Li2016,Zhang2017},
 \begin{equation}
   W(k_f;k_0)=
   \hat{P} \exp[ i \int_{k_0}^{k_f} \hat{A}(k) 
    dk/\hbar],
   \label{k}
 \end{equation}
 where 
 $\hat{P}$ is the path-ordering operator and  
 the matrix representation of the non-Abelian Berry connection $\hat{A}(k)$ is defined as 
 ${A}_{s',s}(k)=\braket{\vec{u}_{s' k}|i \partial_k|\vec{u}_{s k}}$, where $|\vec{u}_{sk}\rangle$ is the periodic Bloch wavefunction in the $s$th band,
 \begin{equation}
 |\vec{u}_{sk}\rangle=e^{isq_Lx}\sum_{j=1}^M\sum_l \tilde{c}^j_{l,s}(k)e^{ilQx}|j\rangle.
 \end{equation}
 We obtain
\begin{widetext}
 \begin{equation}
   A_{s',s}(k)=\sum_j^M \sum_{l,l'}
   \left(\tilde{c}^{j*}_{l',s'}(k) \partial_k\tilde{c}^j_{l,s}(k)\right) \int_0^{2\pi/q_L} e^{i [(s-s') q_L +(l-l')Q]x}dx \sim \delta((s-s') q_L+(l-l')Q).
   \label{<+label+>}
 \end{equation}
 \end{widetext}
 As $|s-s'| q_L \le (n-1) q_L$ and $|l-l'|Q\ge n q_L$ (if $l\neq l'$), to ensure that  $(s-s') q_L +(l-l')Q=0$ is valid, $s=s'$ and $l=l'$ must be satisfied. Thus, 
 $A_{s,s'}(k)\sim \delta(s-s');$ Here, $n=Q/q_L$ is defined
 in the main text.   This means that during the time evolution, there is no transition between bands with different $s$ (different colored bands in Fig.~2 of the main text).

\section{Interaction effects on the expansion of a Gaussian wave packet in the real space}
\begin{figure} [b]
  \includegraphics[width=\linewidth]{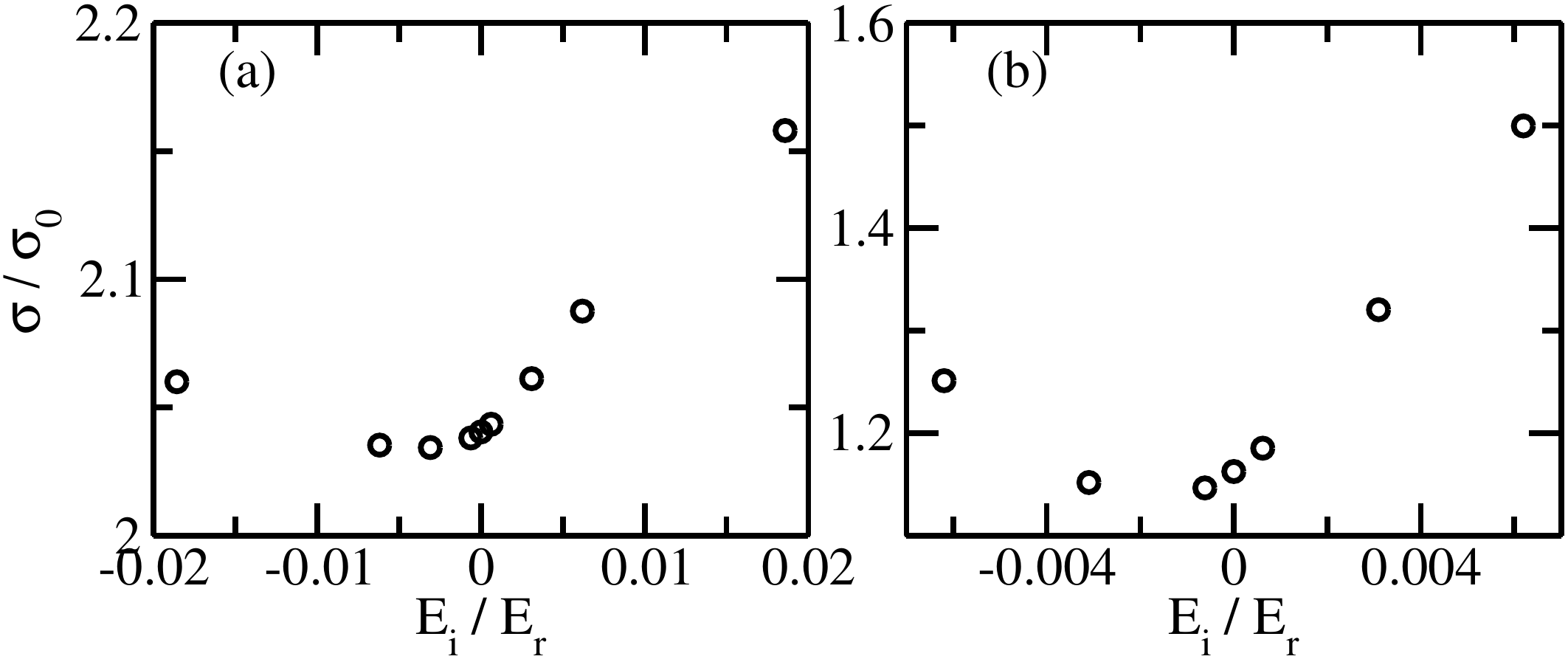}
\caption{
Density spread as a function of initial interaction energy $E_i$ for
$\Omega'=0$ (a) and $\Omega'=2E_r$(b).
}
\label{figs1}
\end{figure}

As discussed in the main text, we 
consider
a 
Gaussian wave packet as the initial state. All the spin components have equal amplitude and phase.
 The width of the wave packet, 
  $\sigma_0=\sqrt{<x^2>-<x>^2}$, 
  is $455/q$.
    We use $\gamma_8=13/21$ to approximate $\gamma$.
The density spread $\sigma_0$ is roughly 3.5 times the underlying period of the Hamiltonian.
Figure~\ref{figs1} 
shows the ratio of the width in a later time $t$ to $\sigma_0$, where 
$t=1000 \hbar/E_r$, as a function of the initial interaction
energy $E_i= \int g \rho_i^2 \text{d} x$ for $\Omega'=0$ and $\Omega'=2E_r$. 
In both cases, weak attractive (repulsive) interactions decrease (increase) the width, $\sigma(t)$. When interactions are strong enough, the width always increase with increasing the strengths of interactions due to the mixing with extended states at high energies.

\section{Expansion dynamics of the ground state in a harmonic trap}
We 
consider weakly interacting bosons in a 
harmonic trap with the trapping frequency $\omega=0.034 E_r /\hbar$, 
and use the 
imaginary time propagation to find the ground state.
For non-interacting systems, the 
ground state has a
Gaussian profile. 
In the presence of 
a
weak repulsive interaction, 
the density is featured with a
Thomas-Fermi
profile.

\begin{figure} [t]
  \includegraphics[width=\linewidth]{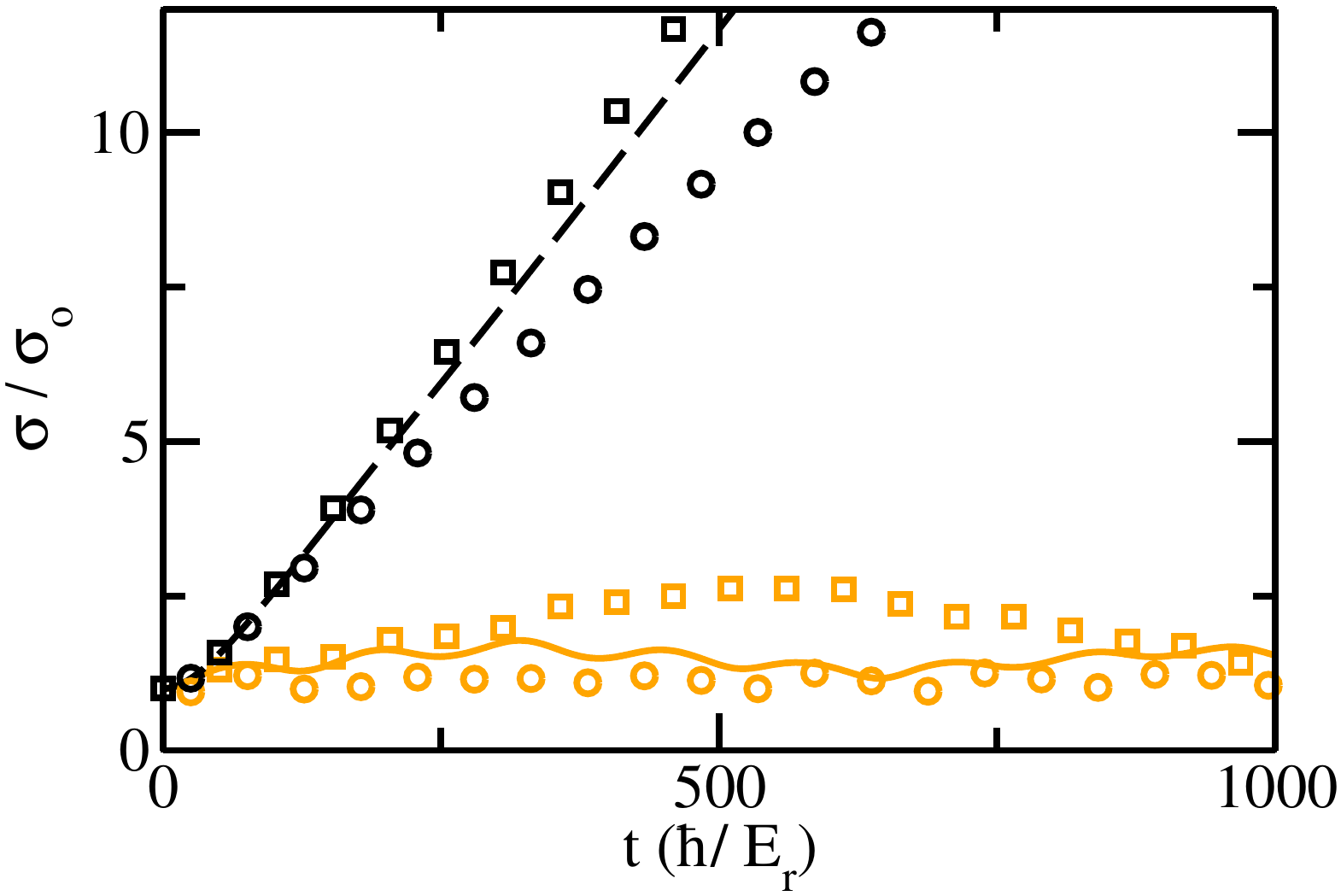}
\caption{
  The evolution of 
  the wavepacket 
after releasing from the trap. 
   The solid orange (black dashed) curve represents results for $\Omega'=2E_r$ ($\Omega'=0$) 
  when the interaction is also turned off in the expansion.  
   Squares and circles represent results for a small repulsion and a small attractive
   interaction 
   added in the expansion, respectively. The absolute interaction strength, $|g|$, is the
   same, 
   and the interaction energies $E_i=\int g \rho_i^2 \text{d} x$ differ because 
   of different densities. 
   For $\Omega'=2E_r$, $E_i/ E_r\approx0.010$ and $-0.039$ for positive and negative $g$, respectively.
   For $\Omega'=0$, $E_i/ E_r\approx0.0098$ and $-0.013$ for positive and negative $g$, respectively.
}
\label{figs2}
\end{figure}

Then we turn off the trap and the interactions, and let the cloud expand.  Fig.~\ref{figs2} 
shows 
the width of
 the wavepacket as a
function of time.
When $\Omega'=0$, the ground state in the trap is an extended one. Therefore, the wavepacket expands quickly after the trap and interactions are turned off. In contrast, when $\Omega'=2E_r$, the ground state is a localized one. Even after turning off the trap and interaction, the width of the wavepacket has little changes, well reflecting the localized nature of the ground state in the trap.  As a further comparison, we turn on a weak interaction in the expansion dynamics once the trap is turned off. It is clear that 
a weak repulsive interaction speeds up the expansion
and a weak attractive interaction slows  it down.

All these results are similar to those presented in the main text. It means that a shallow trap used in typical cold atom experiments do not change our qualitative results and main conclusions. In a shallow trap, though the momentum is no longer a good quantum number and an extended state does not extend to infinity, it can still spread over the whole trap. In contrast, localized states still remain localized, provided that the trapping potential is much smaller than the energy gap between the localized states and delocalized ones.

\section{
A realistic setup 
in experiments}
\begin{figure} [t]
  \includegraphics[width=\linewidth]{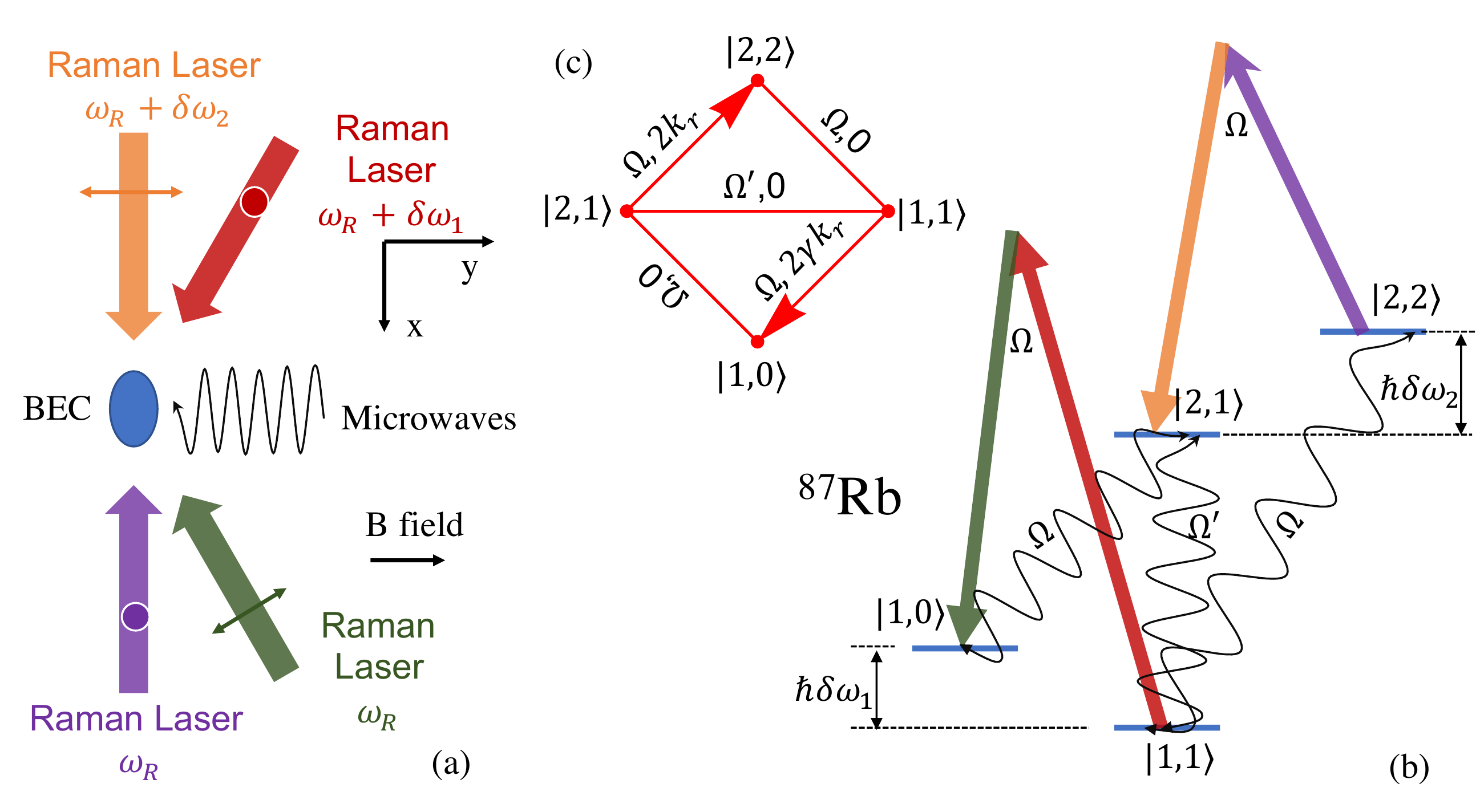}
\caption{
  (a) 
  A Bose Einstein Condensate is 
  confined in a harmonic trap. 
  Straight arrows represent Raman lasers and wiggles indicate microwaves.
  (b) A 
  scheme to couple hyperfine states of $^{87}$Rb. Four states are
  cyclically coupled by 
  pairs of Raman lasers (arrows) and microwaves (wiggles). 
  In addition, a microwave couples states $\ket{1,1}$ and $\ket{2,1}$
  with coupling strength $\Omega'$.
  (c) A 
  schematic of the cross sections of two cylinders 
  or tori when they are glued together.
  The coupling strengths 
  and momenta 
  transferred are labeled.
}
\label{figs3}
\end{figure}

Our scheme can be realized in laboratories as a simple generalization of some current experiments~\cite{Li2018}.  
We choose four hyperfine states of $^{87}$Rb,
$\ket{F,m_F}=\ket{2,2},\ket{2,1},\ket{1,0},\ket{1,1}$.
States $\ket{2,2}$ and $\ket{2,1}$ are resonantly coupled 
by one pair of counter
propagating Raman beams with recoil momentum $k_r=2\pi/\lambda$,
where $\lambda=790nm$ is the wavelength. The momentum transferred is $2k_r$.
States $\ket{1,0}$ and $\ket{1,1}$ are resonantly coupled 
by another pair of
Raman beams with the same wavelength that is tilted with a small angle $\theta$ with respect to the 
first pair of
Raman beams. The momentum transferred is $2k_r \cos\theta$, where the angle
$\theta$ is
chosen such that $\cos\theta$ is irrational or rational, say, 
$\cos\theta=\gamma_i$, where $\gamma_i$ is defined in the main text. 
One microwave 
couples states $\ket{2,2}$ and $\ket{1,1}$, 
and another
couples 
states $\ket{2,1}$ and $\ket{1,0}$.
This completes a cyclic coupling between four states and delivers a synthetic cylinder. Replacing Raman beams by LG beams, this creates a synthetic torus. 
All 
coupling strengths 
are scaled with
$2E_r=\hbar^2 (2k_r)^2/m$, where $m$ is the mass of a Rubidium atom.
As discussed in the main text, 
eigenstates 
are extended, even when $\cos\theta$ is irrational and the Hamiltonian is quasi-periodic.  

To glue two cylinders or tori, 
an extra microwave coupling between states
$\ket{2,1}$ and $\ket{1,1}$ could be added. The coupling strength is $\Omega'$.
The experimental setup is depicted in Fig.~\ref{figs3}.

Using realistic experimental parameters, we perform 
time-dependent simulations using the Gross-Pitaevskii
equation. We 
consider 15000 $^{87}$Rb atoms 
in a three-dimensional harmonic trap with an angular
trapping frequency $2\pi \times 500$Hz. 
We compute 
the ground states in the presence of Raman and microwave couplings, interactions and the harmonic trap. 
After the initial state is prepared, we turn off the trap along 
$\hat{x}$, 
the direction of momentum transfer. 
The 
transverse 
confinement remains.  
All the inter- and intra-species scattering lengths are well approximated by 
93$a_0$, where $a_0$ is the Bohr radius. The angle is chosen such that $\cos
\theta=13/21$.

\begin{figure} [htb]
  \centering
  \includegraphics[width=\linewidth]{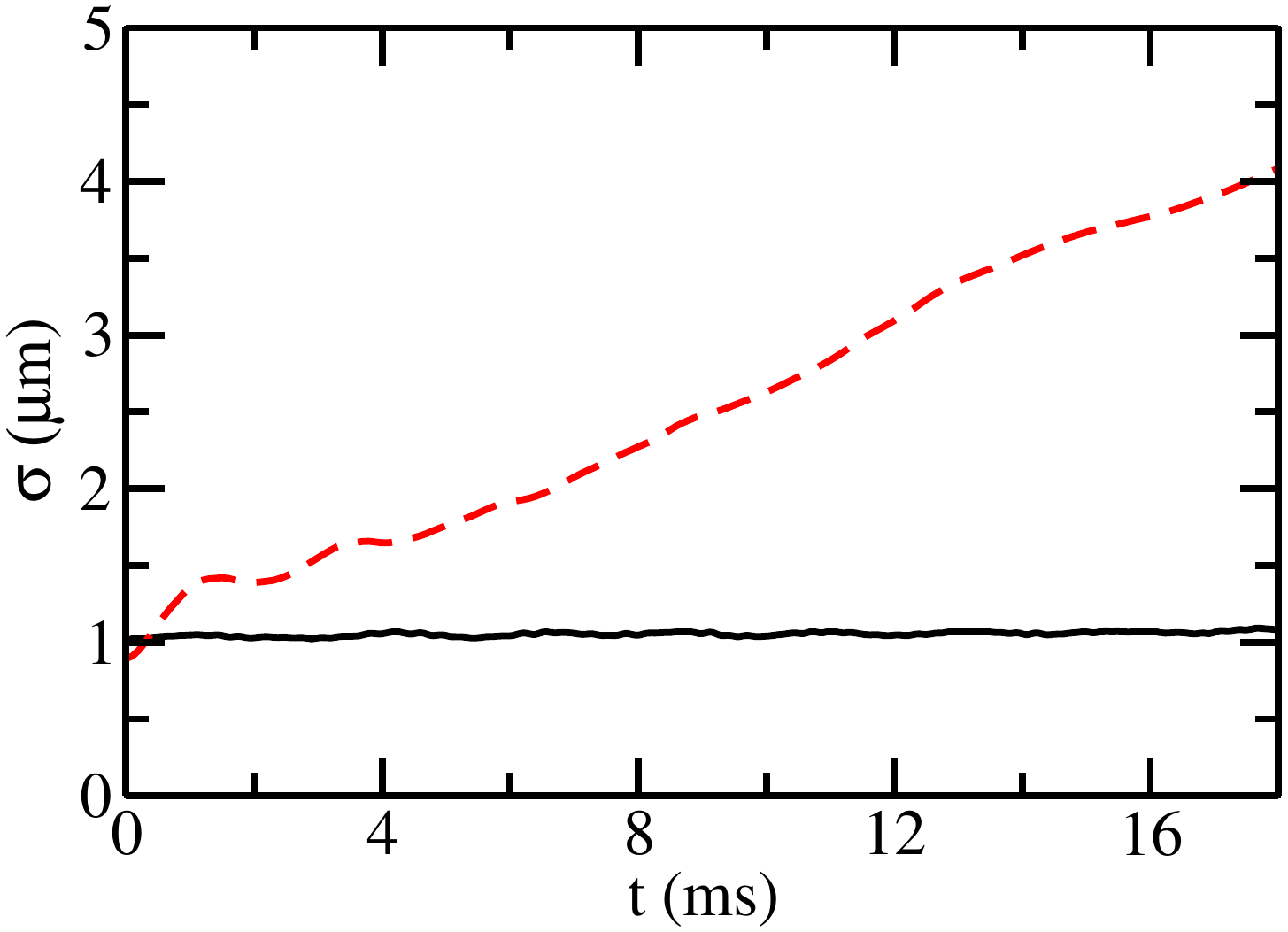}
\caption{
  Expansion dynamics when the ground state 
  is released from the trap 
  along the
  $x$ direction. 
  Solid line and dashed lines 
  represent results of $\Omega'=2E_r$ and $\Omega'=0$, respectively. 
}
\label{figs4}
\end{figure}

Fig.~\ref{figs4} 
shows the width of the wavepacket, $\sigma$, as a
function of time $t$ 
after the trap and the interaction are turned off.
The wavepacket expands fast when $\Omega'=0$, as a consequence of the extended ground state in the trap. 
It 
expands little when
$\Omega'=2E_r$, reflecting the localized nature of the initial state in the trap after glueing two cylinders.

\section{
Comparisons between the uniform and nonuniform flux}

\begin{figure} []
  \includegraphics[width=\linewidth]{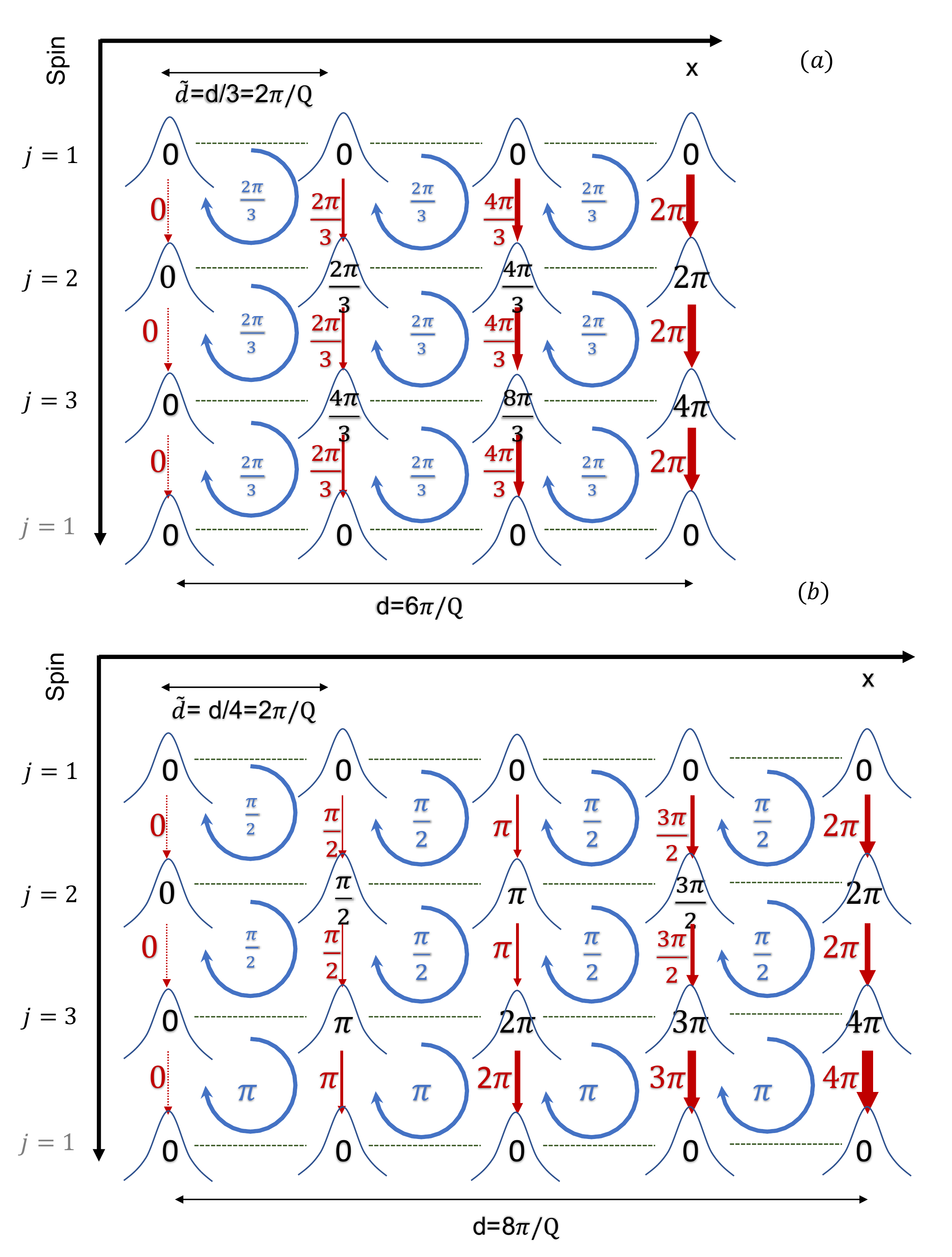}
\caption{ Curves represent the density profiles of each individual spin component near the maxima. The phases of the wavefunctions are also shown. Red arrows represent the spatially dependent coupling along the synthetic direction. Maxima of the density form plaquettes, in which  the flux per plaquette are shown. 
}
\label{figs5}
\end{figure}

The ground states of the systems for Fig.~2(a) and Fig.~2(b) in the
main text are 
depicted
in Fig.~\ref{figs5}(a).
The 
density modulation of each spin component is 
determined by the total momentum 
$Q$, or equivalently, the total flux penetrating a unit length.
The period of the density oscillation $\tilde{d}=2\pi/Q$.
However, the relative phases between different spin components depend on how the flux is distributed on the surface. 
For the system in Fig.~2(a), $n=3$, $q_L=q=Q/3$, and the flux is uniformly distributed. The relative phase has a period of  $d=2\pi/q=3\tilde{d}$.
The lattice spacing is therefore $d=6\pi/Q$. 
In contrast, for the system in Fig.~2(b), $n=4$, $q_L=3q/4$, and the distribution of the flux is nonuniform.   The relative phase has a period of $8\pi/Q$, and 
the lattice spacing is therefore $8\pi/Q$.

\begin{thebibliography}{47}
\bibitem [{not({\natexlab{a}})}]{note1}%
  \BibitemOpen
  \bibinfo {note} {We normalize the wave function according to
  $\int_0^{2\pi}|\vec{\psi}(\phi)|^2 d \psi=2\pi$
  ($\int_0^{L}|\vec{\psi}(\phi)|^2 d x=L$) for the torus (cylinder), where $L$
  should be multiples of the period of the underlying Hamiltonian. Thus, the
  wave function and the density is dimensionless in both cases.}\BibitemShut
  {Stop}%
\bibitem [{not({\natexlab{c}})}]{note3}%
  \BibitemOpen
  \bibinfo {note} {Only half of the period is shown. Due to the inversion
  symmetry and nonsymmorphic symmetry, there are, in general, 2 or 4 peaks in a
  period, $2\pi a_8/q$. As the approximation order increase, the decrease of
  the width of the wavefunction is accompanied by the increase of the period
  and the separation between peaks.}\BibitemShut {Stop}%
\end{thebibliography}
\end{document}